# Optical Depth and Vertical Profile of Stratospheric Aerosol based on Multi-Year Polarization Measurements of the Twilight Sky

Ugolnikov O.S.[1], Maslov I.A.[1,2]

[1]Space Research Institute, Russian Academy of Sciences,
84/32 Profsoyuznaya st., Moscow, 117997, Russia
[2]Moscow State University, Sternberg Astronomical Institute,
13 Universitetsky pr., Moscow, 119234, Russia

E-mail: ougolnikov@gmail.com, imaslov@iki.rssi.ru

The method of detection of light scattering on stratospheric aerosol particles on the twilight sky background is considered. It is based on the data on sky intensity and polarization in the solar vertical at zenith distances of up to 50° from the sunset till the moment of the Sun depression to 8° below the horizon. The measurements conducted in central Russia since 2011 had shown the negative trend of stratospheric aerosol content, this can be related with the relaxation of the stratosphere after the number of volcanic eruptions during the first decade of XXI century.

## 1. Introduction

Stratospheric aerosol is an important component significantly defining chemical processes in the middle atmosphere and ozonosphere (Hofmann and Solomon, 1989) and thermal conditions of the Earth's surface. A multifold increase in the number of sulfate particles in the stratosphere after major volcanic eruptions leads to the albedo growth and global cooling for several years. Present state of stratospheric aerosol study and related problems is completely described by Kremser et al. (2016).

A larger amount of stratospheric aerosol can be the cause of optical events during the early twilight period when the stratosphere is still illuminated by the Sun, while the troposphere is not (Lee and Hernandes-Andres, 2003). Such phenomena mostly manifest themselves as the intensively red or purple color of dusk. It was observed and correctly interpreted in the early XX century by Gruner and Kleinert (1927). Stratospheric aerosol particles are small; their size under background conditions is about 0.1 μm and several times larger after volcanic eruptions (Deshler et al., 2003). According to the Mie theory, the scattering coefficient rises in shorter wavelengths. However, solar emission in the blue spectral range is strongly absorbed, the effective scattering takes place at higher altitudes, while the particle density rapidly decreases above 25 km. In the green and yellow spectral ranges, additional absorption is caused by Chappuis lines of stratospheric ozone. All these effects explain the red and purple color of the twilight glow segment after volcanic eruptions.

Aerosol particles were discovered during the 1961 balloon experiment (Junge et al., 1961). Regular balloon measurements had started in the 1970's (Deshler et al., 2003). That period covered intense eruptions of El Chichon in 1982 and Mt. Pinatubo in 1991. They significantly changed not only the particles' density but also their microphysical properties (Bauman et al., 2003). The stratosphere remains disturbed for about five years after an eruption. This helped to study post-volcanic aerosol but hindered the research of background (undisturbed) aerosol due to the episodical nature of such circumstances.

A long volcanically quiescent period started in the late XX century, during which there has been a number of weaker eruptions, making it possible to arrange long series of background stratospheric aerosol measurements by the same technique intended to identify possible long-term trends. Balloon measurements (Hofmann and Rosen, 1980) showed an increase in the particle number compared with initial results of Junge et al. (1961). However, it could be associated with a better sensitivity of the experiment. A long series of measurements in Laramie, Wyoming, USA (Deshler et al., 2006) after excluding the post-volcanic periods did not show any significant trends of the aerosol content growth. On the opposite, an insignificant negative trend was observed.

However, a several percent increase of particle density every year was detected in the early XXI century (Solomon et al., 2011; Liu et al., 2012). This issue is particularly important, since background aerosol can basically originate from the emissions of carbonyl sulfide OCS (Crutzen, 1976). These emissions can be partially anthropogenic (Campbell et al., 2015), and the mixing ratios of OCS are above the pre-industrial values (Aydin et al., 2014). Another source of sulfur particles is sulfur dioxide $SO_2$ (Brock et al., 1995). It should be noted that the light scattering function of tiny particles is closer to symmetric and such particles can significantly increase the Earth's albedo and decrease the surface temperature (Hinds et al., 1999).

The uncertainty of trend values and its connection with ecological problems justify the need for its estimation using various methods of systematical aerosol monitoring. Optical phenomena during twilight, caused by stratospheric aerosol and often observed by occasion, indicate the efficiency of analyzing scattered radiation for the analysis of the integral values and vertical profiles of aerosol. Measuring the sky background polarization broadens the scope and accuracy of this method, since the polarization properties of aerosol scattering, defined by the Mie theory, differ from molecular and multiple scattering.

Three-color measurements of the sky background conducted during the spring and early summer of 2016 showed an excess of brightness and decrease of polarization of the dusk sky segment during the light stage of twilight (solar zenith angle 92-95°) in the red spectral range. These effects were interpreted by Ugolnikov and Maslov (2018ab) as the presence of stratospheric aerosol scattering. The particles' size distribution was estimated by two independent methods and found to be close to the typical multi-year distribution established as a result of the balloon experiments (Deshler at al., 2003, 2006). The mean particle radius reached the maximum (about 0.1 µm) in the Junge layer, slightly above 20 km, that was also in good agreement with previous estimations.

However, the vertical profile of particle density was not clearly defined, since effective scattering takes place in a wide range of altitudes in the atmosphere at any moment of twilight, and profile retrieval is a difficult inverse problem. Its solution is possible if some a priori information about aerosol is known. Three-color measurements were conducted during just several months, trend estimation was not the goal of that work. Here, we consider the procedure of vertical profile retrieval and trend estimation for stratospheric aerosol based on one-color polarization measurements of the twilight sky background for the last eight years.

## 2. Observations

The experiment was performed with a Wide-Angle Polarization Camera (WAPC, Ugolnikov and Maslov, 2013ab) installed in Chepelevo (55.2°N, 37.5°E), Moscow region, Russia. The camera worked in summer period since 2011 and consisted of three consecutive lenses and a rotating polarization filter. Its angular field was about 140°. In this research, we used the data obtained in solar vertical with zenith distances $\zeta$ up to 50° at 5° intervals. The value of $\zeta$ is positive in dusk area and negative in the solar vertical part opposite to the Sun. The value of polarization measured in these sky points is naturally the second normalized Stokes component which is positive if the polarization directed perpendicular to the solar vertical and negative in the case of polarization direction parallel to the solar vertical.

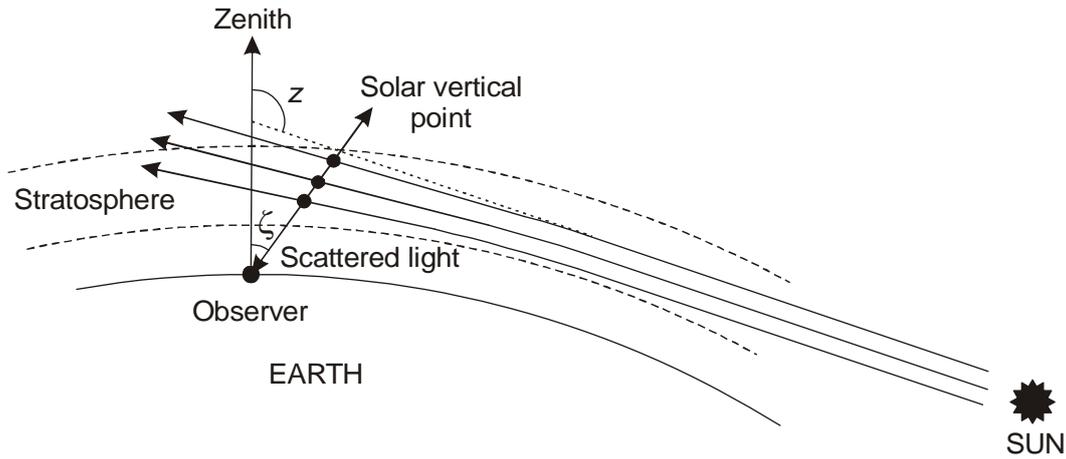

*Figure 1. Single scattering geometry during the twilight.*

The sky background was measured in a wide spectral band with effective wavelength of 540 nm and FWHM of 90 nm. This band is close to the G channel of RGB-observations with this camera in 2016; those were also included in this work. The measurements started before the sunset, continue during the whole twilight and night, and finish after the sunrise. The analysis made here is based on measurements of the solar zenith angles $z$ from 90° to 98°. The camera position, field curvature, and flat field are defined basing on star image astrometry and photometry during the night. The geometrical picture of single scattering with angle definition is shown in Figure 1.

Figure 2 shows the dependency of the logarithm of intensity ratio in symmetric points of the solar vertical during the evening twilight of August 28, 2016 (solid lines). This dependency, together with the sky background polarization shown in Figure 3 for the same evening, reflects the changes in background composition during the twilight. In the dark stage, at solar zenith angle 98-99°, the asymmetry factor and polarization significantly decrease. As it was shown by Ugolnikov (1999), Ugolnikov and Maslov (2002, 2007), multiple scattering prevails during this stage. Its polarization is practically symmetric in the solar vertical during dark twilight (Ugolnikov and Maslov, 2013b).

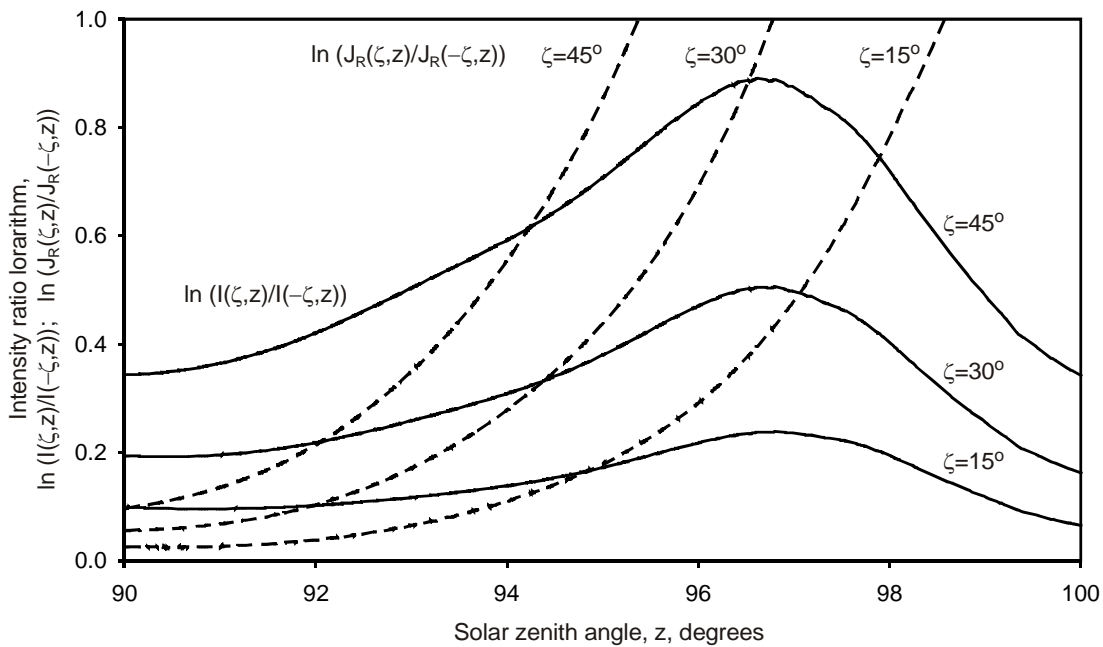

*Figure 2. Logarithm of intensity ratio in symmetric points of solar vertical for total background (measurements, solid) and single molecular scattering (model, dashed), evening twilight of August, 28, 2016.*

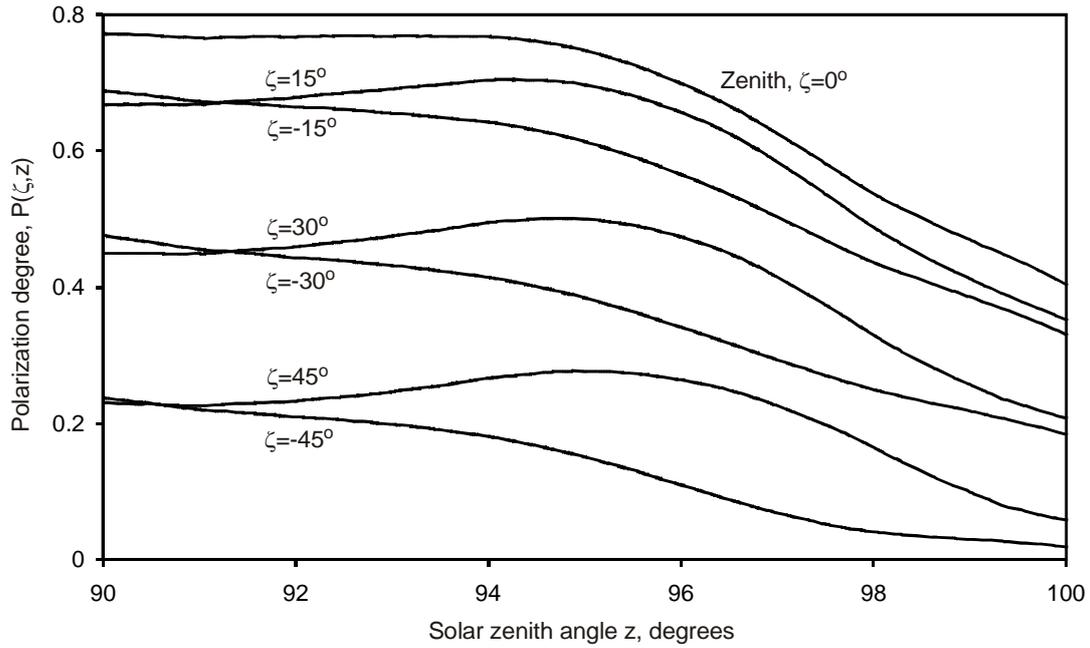

*Figure 3. Polarization degree of twilight sky background in different points of solar vertical, the same twilight as in Figure 2.*

Single scattering contribution increases at the lighter stage of twilight. Owing to a significant difference in the Earth shadow altitude, single scattering intensity is highly asymmetric (the model brightness ratio is shown with dashed lines in Figure 2), and this background component appears first in the dusk segment of the sky, where the shadow is lower. This effect leads to excessive brightness and polarization in the dusk area, which reach their maximum at the solar zenith angle of about 96°. Polarization of the sky background near the zenith is as high as 0.8 in good atmospheric conditions, which is due to the Rayleigh scattering that is almost linearly polarized at 90° from the Sun.

During the even lighter twilight stage, the Earth's shadow becomes almost horizontal and asymmetry effects described above decrease. When the Sun is close to the horizon, opposite trends of polarization are seen on both sides from the zenith, which is due to a change in the single scattering angle (Ugolnikov, 1999). It is worthy of note that polarization becomes symmetric when the Sun is still below the horizon. At the moment of sunrise, polarization in the dusk segment is less than in the opposite part of the sky. These effects are the qualitative manifestation of the stratospheric aerosol presence. The assay procedure is described in the next chapter of this article.

**3. Detection of aerosol scattering**

As we spoke above, retrieval of the vertical profile of a certain scattering component based on the twilight sky photometry and polarimetry is a complicated inverse problem due to the significant thickness of the effective scattering layer in the atmosphere during a certain twilight stage. It can be solved with model assumptions restricting the vertical resolution of the analysis.

In the model used herein, stratospheric aerosol is a mixture of two fractions. The first one is represented by tiny particles with a size significantly less than the wavelength. Their scattering properties are similar to molecular scattering. The second component has lognormal size distribution of particles with median radius $r$=0.08 μm and distribution width $\sigma$=1.6. This distribution characterizes aerosol particles according to balloon measurements (Deshler et al., 2003) and is confirmed by a color analysis of the twilight sky (Ugolnikov and Maslov, 2018ab).

Following this paper, we take the value of refracting index of sulfur aerosol particles equal to 1.44. Figure 4 shows angular dependencies of scattered light intensity and polarization for both components (intensity is calculated for the unity absorption coefficient). It is obvious that brightness excess and depolarization in the dusk area considered in (Ugolnikov and Maslov, 2018ab) are associated with a large aerosol fraction.

Introduce the array of altitude values $h_i$, separated by equal intervals $\Delta h$=5 km. Assume $A_{1i}$ ($A_{2i}$) is the ratio of the extinction coefficient of the first (second) aerosol fraction and Rayleigh scattering at altitude $h_i$. The vertical profile of this ratio is represented as a broken line with knots at altitudes $h_i$. For altitude $h$ between knots $i$ and $(i+1)$, the extinction coefficients ratio is:

$$a_{1,2}(h) = A_{1,2}(h_i)\frac{h_{i+1} - h}{\Delta h} + A_{1,2}(h_{i+1})\frac{h - h_i}{\Delta h}. \qquad (1)$$

The coefficient of aerosol extinction is found by multiplying $a(h)$ by Rayleigh extinction coefficient $R(h)$. Rayleigh coefficient $R(h)$ and the single scattering field are defined by the profiles of temperature and atmospheric ozone that significantly absorbs emission in the instrumental band. These profiles are taken for each observation date from the EOS Aura/MLS satellite data (EOS MLS Science Team, 2011ab) for nearby locations. This allows finding the dependence of intensity and polarization of Rayleigh scattering $J_{0R}(A_{1,2i}, \zeta, z)$, $p_R(\gamma)$, where $z$ is the solar zenith angle. Since the refraction angle is small, polarization is defined by scattering angle $\gamma = z - \zeta$. Aerosol influences the Rayleigh scattering intensity by absorption of tangent solar emission; therefore, the intensity depends on the sums $(A_1+A_2)_i$ only. Besides, near-ground aerosol layers, not involved into this model, decrease the intensity of all twilight background fractions after the scattering event. This factor is not taken into account while calculating function $J_{0R}$, it is corrected by star photometry on nighttime images and clarified during a latter analysis.

The model with values $(A_1 + A_2) = 1$ below 10 km and $(A_1 + A_2) = 0$ above 10 km is taken as zero-order approximation. The relation between intensity of single Rayleigh scattering $J_R$ fixed by the camera and model value $J_{0R}$ is as follows:

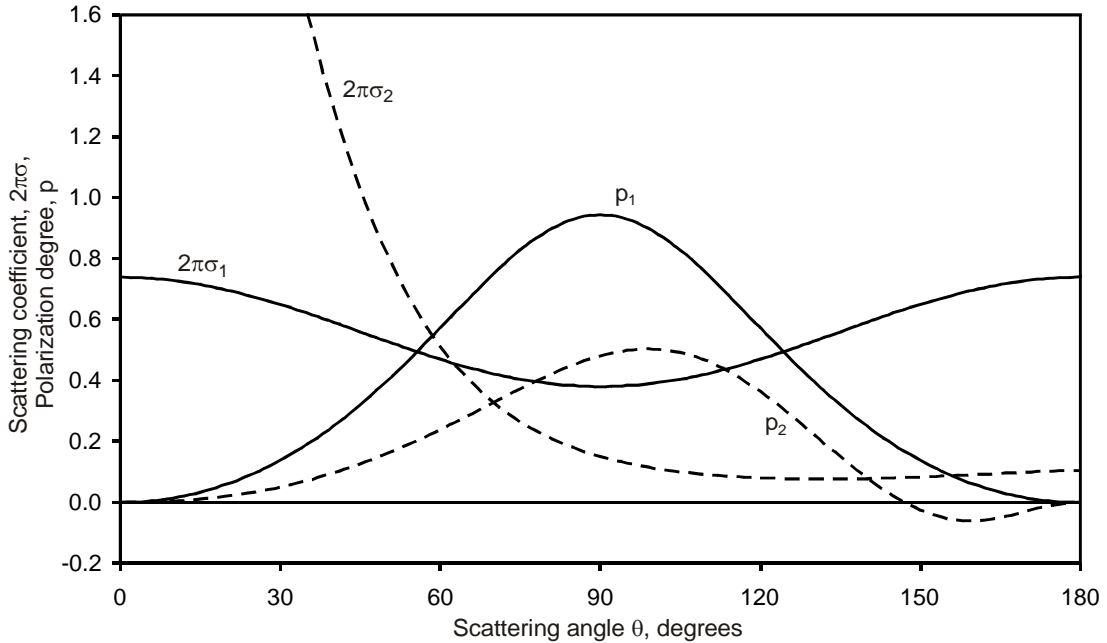

*Figure 4. Scattering coefficient (per unit of extinction) and polarization degree for molecular scattering (solid lines) and large fraction of aerosol (dashed). Negative values of polarization correspond to its direction in the scattering plane.*

$$J_R(\zeta, z) = J_{0R}(A_{1,2i}, \zeta, z) \cdot (K_1 + K_2\zeta^2). \tag{2}$$

In this case, the value of $K_1$ designates the camera sensitivity in the zenith, and term $K_2\zeta^2$ reflects changes in the sensitivity as the camera moves away from the zenith, as well as it reflects light absorption by near-ground aerosol. These factors are initially corrected by star photometry, expression (2) is used for their further refinement. Simple expression (2) is good enough for zenith angles $\zeta<50°$ used in this work.

Denote the general intensity and polarization of the sky background as $I(\zeta, z)$ and $P(\zeta, z)$. Intensity and polarization of background fraction without single Rayleigh scattering are expressed as follows:

$$j(\zeta, z) = I(\zeta, z) - J_R(\zeta, z); \quad q(\zeta, z) = \frac{I(\zeta, z)P(\zeta, z) - J_R(\zeta, z)p_R(\zeta, z)}{I(\zeta, z) - J_R(\zeta, z)}. \tag{3}$$

Consider the twilight period corresponding to the solar zenith angles from 96° to 98°. Due to Chappuis absorption bands of stratospheric ozone, the effective shadow of the Earth is 15-20 km higher than its geometric altitude, and effective single scattering during this period takes place above the stratospheric aerosol layer. Assume that the sky background consists of Rayleigh and multiple scattering. Polarization of the latter is symmetric with respect to the zenith. The values of $K_1$ and $K_2$ can be found from equations (2-3) by the following criterion:

$$\sum_z \sum_\zeta (q(\zeta, z) - q(-\zeta, z))^2 = \min. \tag{4}$$

The squares of residuals are summed at different sky points with zenith angles $\zeta$ up to 50° and twilight moments with solar zenith angles $\zeta$ from 96° to 98°. Calculating the values of $K_1$ and $K_2$, we can find molecular scattering intensity $J_R(\zeta, z)$ for any twilight stages. Subtracting it from the total measured intensity, we find the dependencies of $j(\zeta, z)$ and $q(\zeta, z)$ for the entire twilight period. During its light stage, this fraction consists of molecular scattering (designated by $j_M(\zeta, z)$ and $q_M(\zeta, z)$) and single aerosol scattering ($J_A(\zeta, z)$, $p_A(\zeta, z)$).

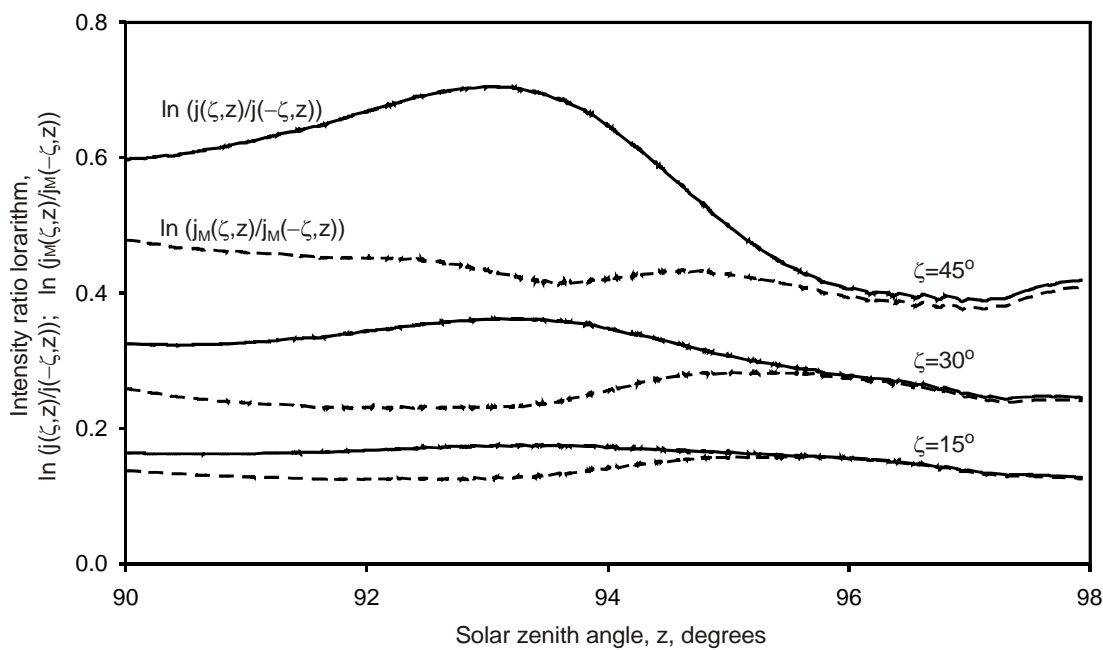

*Figure 5. Logarithm of intensity ratio in symmetric points of solar vertical after subtraction of single molecular scattering (solid lines) and after total subtraction of single scattering (dashed lines), evening twilight of August, 28, 2016.*

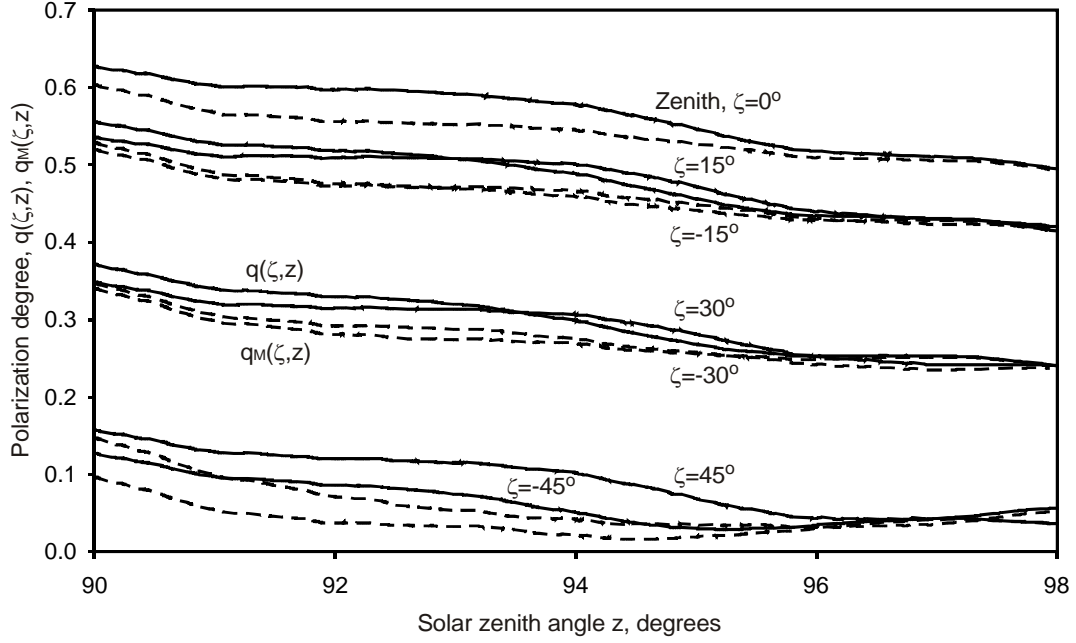

*Figure 6. Polarization degree of sky background in solar vertical points after subtraction of single molecular scattering (solid lines) and after total subtraction of single scattering (dashed lines), the same twilight as in Figure 5.*

Figures 5 and 6 show the dependencies of the logarithm of brightness ratio in symmetric points of the solar vertical $\ln(j(\zeta, z)/j(-\zeta, z))$ and polarization $q(\zeta, z)$ after subtracting Rayleigh scattering (solid lines). The values are similar to the total background characteristics shown in Figures 2 and 3 for the same evening twilight of August 28, 2016. Naturally, the asymmetry of brightness and polarization is weaker than in Figures 2, 3; however, these effects are also seen. Polarization in the dusk segment is slightly higher than in the opposite part of the sky at solar zenith angles 94°-96°, which corresponds to scattering altitudes of 40-50 km. This effect can be caused by tiny aerosol particles having the same polarization properties as Rayleigh scattering. During the lighter stage, the dusk segment polarization decreases and the background brightness becomes excessive at the same time. It is the effect of a large aerosol fraction with asymmetric scattering function and weaker polarization (see Figure 4).

Analyzing the light twilight period, we cannot assume exact symmetry of polarization of multiple scattering in the solar vertical and a constant ratio of intensities in symmetric points. It is due to a change in the effective altitude of multiple scattering and greater inclusion of the tropospheric aerosol scattering into this background component. These processes are gradual and the changes in the multiple scattering characteristics are very slow. Following (Zaginailo, 1993; Ugolnikov, 1999), we can assume differential properties of multiple scattering:

$$\frac{d \ln j_M(\zeta, z)}{dz} = \frac{d \ln j_M(-\zeta, z)}{dz}; \quad \frac{dq_M(\zeta, z)}{dz} = \frac{dq_M(-\zeta, z)}{dz}; \qquad (5)$$

These multiple scattering characteristics are defined as follows:

$$j_M(\zeta, z) = I(\zeta, z) - J_R(\zeta, z) - J_A(\zeta, z);$$
$$q_M(\zeta, z) = \frac{I(\zeta, z)P(\zeta, z) - J_R(\zeta, z)p_R(\zeta, z) - J_A(\zeta, z)p_A(\zeta, z)}{I(\zeta, z) - J_R(\zeta, z) - J_A(\zeta, z)}. \qquad (6)$$

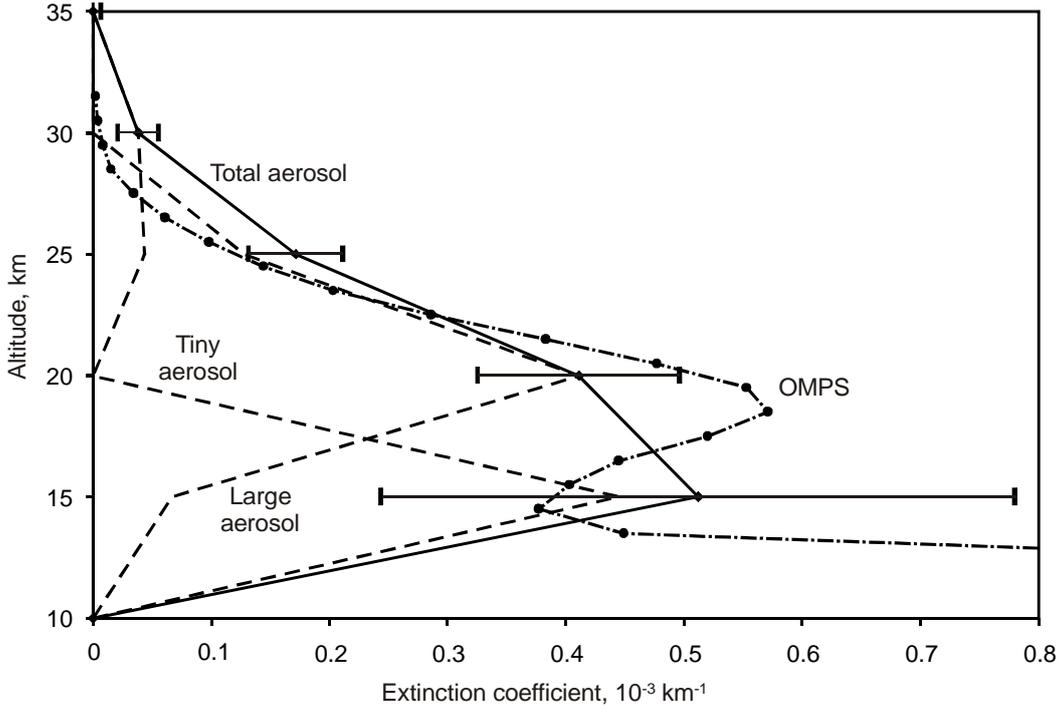

*Figure 7. Altitude profile of stratospheric aerosol (tiny, large and total) in the evening of August, 28, 2016, compared with OMPS satellite profile at 525 nm.*

The properties of molecular scattering $J_R$, $p_R$ are known from equation (2) and Rayleigh scattering function; the characteristics of aerosol scattering $J_A$, $p_A$ are defined by Mie theory and the array of parameters $A_{1i}$, $A_{2i}$ (1). The criterion of their optimal values is the minimum of square residuals summed by sky points and twilight moments (5):

$$\sum_z \sum_\zeta \left( \frac{d \ln j_M(\zeta, z)}{dz} - \frac{d \ln j_M(-\zeta, z)}{dz} \right)^2 + \left( \frac{dq_M(\zeta, z)}{dz} - \frac{dq_M(-\zeta, z)}{dz} \right)^2 = \min.; \quad ; \qquad (7)$$

This nonlinear least-squares problem is solved with additional restrictions: $A_{1i} \geq 0$, $A_{2i} \geq 0$. The measurements at solar zenith angles from 90° to 98° and point zenith angles of up to 50° are taken into account. The obtained values of $A_{1i}$ and $A_{2i}$ are used to correct the light extinction in molecular scattering model $J_{0R}(A_{1,2i}, \zeta, z)$, and then the iteration process is repeated until its convergence.

An example of retrieved profile of stratospheric aerosol (small and large fractions and total profile) for the evening twilight on August 28, 2016 is shown in Figure 7. For a comparison, the aerosol profile for a close location on the same day was plotted based on satellite measurements (OMPS, DeLand, 2016). The agreement between the profiles is quite good; however, the twilight data has a lower altitude resolution, the causes were described above. The Junge layer (at the altitude of about 20 km) is characterized by a maximum density of the large aerosol fraction; this was also noted in (Ugolnikov and Maslov, 2018ab).

Figure 8 shows the dependencies of contribution of single molecular and aerosol scattering in the total twilight background on the solar zenith angle for different solar vertical points. Subtracting aerosol scattering intensity $J_A$ in equation (6), we find the multiple scattering background intensity $j_M$ and polarization $q_M$. Dashed lines in Figures 5 and 6 show the dependencies of the logarithm of the ratio of brightness in symmetrical points of the solar vertical, $\ln (j_M(\zeta, z)/j_M(-\zeta, z))$, and polarization $q_M(\zeta, z)$. As assumed, these values change slowly during twilight, polarization is almost symmetric with respect to the zenith.

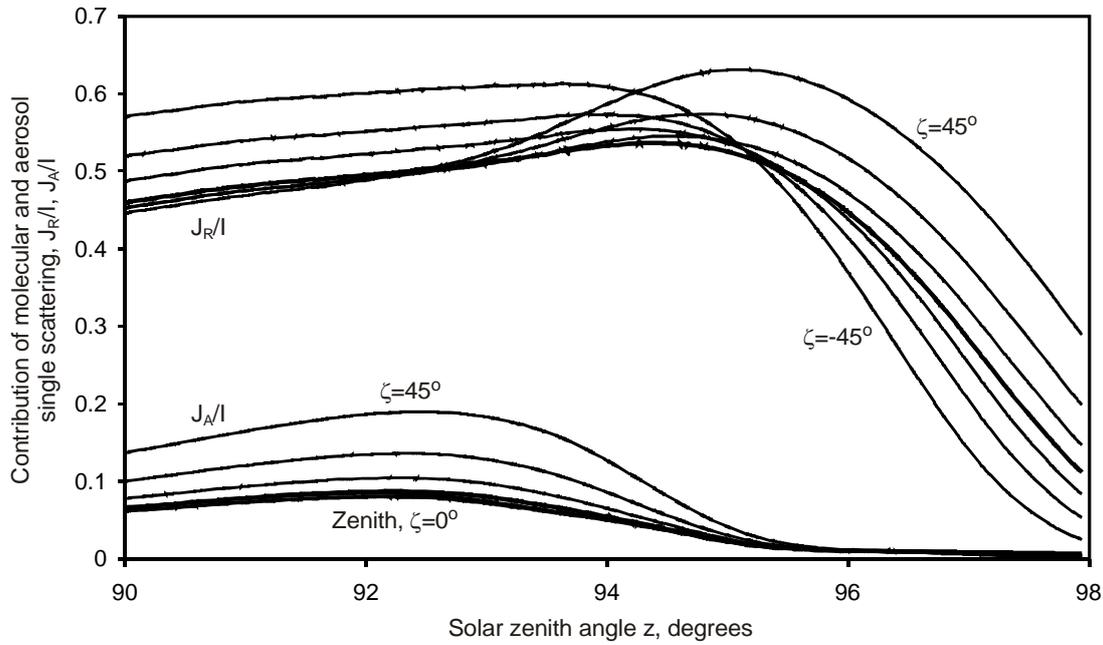

*Figure 8. Contribution of molecular and aerosol single scattering in the total twilight background in different solar vertical points, the same twilight as in Figures 5-7.*

Based on the above procedure for observed data obtained in 2011–2018, we can find the dependence of the optical depth of stratospheric aerosol (above 15 km) on time. This dependence is shown for a large aerosol fraction in Figure 9 and for the total amount in Figure 10. It is important to note that the accuracy of single measurement is low (the points with errors less than 0.003 are plotted), but a general review does not show any positive trend. Moreover, there is a decrease in stratospheric aerosol. Typical values of optical depth in the initial years (about 0.005-0.006 for 540 nm) are close to the data presented by Solomon et al. (2011) for the same period and close wavelength. This value decreases almost twice in recent years, being close to post-Pinatubo epoch in 1998-2002 (Vernier et al., 2011). We also note that decrease in background stratospheric aerosol after 2010–2011 was also noted during the balloon, lidar and satellite measurements (Ridley et al., 2014; Kremser et al., 2016).

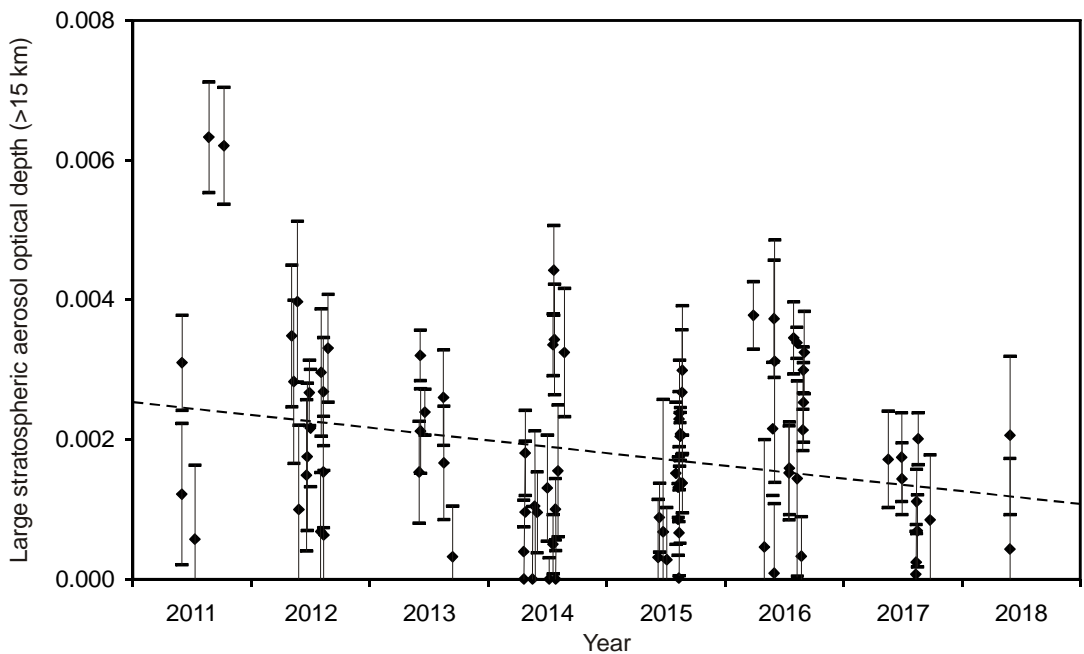

*Figure 9. Optical depth of large fraction of stratospheric aerosol above 15 km by twilight analysis in 2011-2018.*

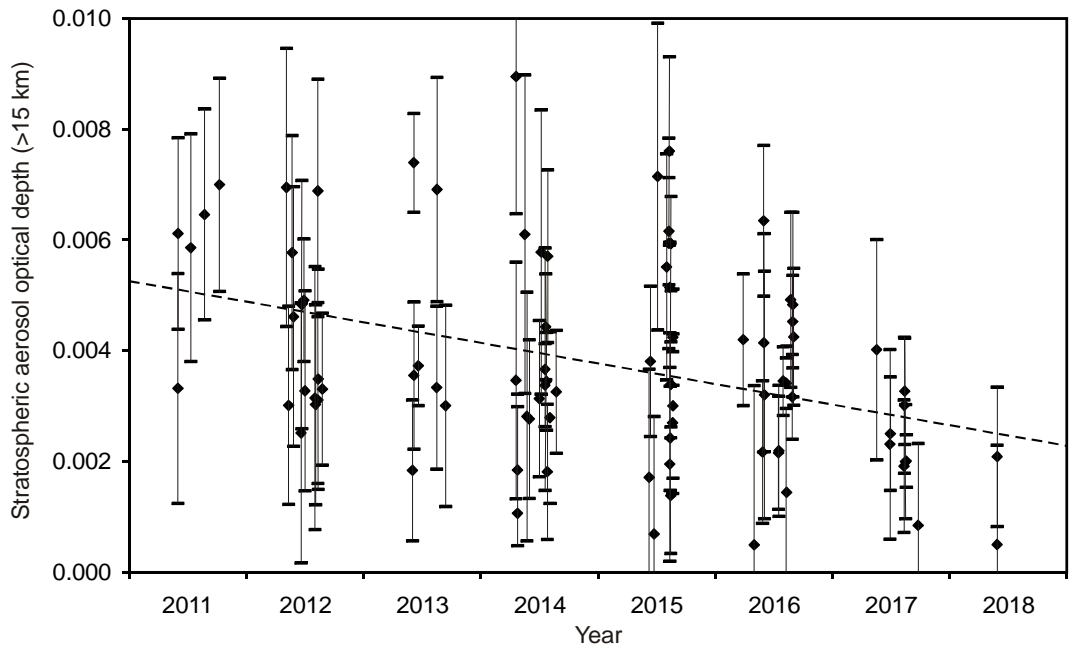

*Figure 10. Total optical depth of stratospheric aerosol above 15 km by twilight analysis in 2011-2018.*

## 4. Discussion and conclusion

The search of long-term trends of different parameters of the stratosphere and mesosphere and their relationship with anthropogenic factors is one of the basic problems of the contemporary research of the middle atmosphere. More or less exact measurements of these values became possible just in the last decades; the consequences of anthropogenic influence on the middle atmosphere are not clear yet.

Signs of possible trend of stratospheric aerosol optical depth have been detected several times since the last decades of the XX century; however, some long-term series of measurements pointed to its absence. The basic cause of uncertainties was the volcanic activity with two major eruptions, El Chichon and Mt. Pinatubo, those changed the optical properties of the entire stratosphere and made the background aerosol study impossible.

The beginning of XXI century is considered as volcanically quiescent; however, effects of some eruptions were optically noticeable in the stratosphere: Tavurvur in 2006 (caused depolarization of the twilight sky fixed by Ugolnikov and Maslov (2009)), Kasatochi in 2008, Sarychev in 2009, Eyjafjallajökull in 2010. As it was shown by Vernier et al. (2011), basing on multi-technique measurements and confirmed by modeling (Neely et al., 2013), these eruptions were basic reason of positive trend of stratospheric aerosol reported in first decade of XXI century, while anthropogenic $SO_2$ could not play significant role in this process.

These statements could be checked during the following decade, provided it is more volcanically quiescent than the previous one. Decrease of stratospheric aerosol optical depth after 2011 was detected in many experiments (see the review by Kremser et al. (2016)), and here we see that stratospheric aerosol had returned to its level of 2000 that is typical for the non-volcanic state. Results of this study indicate the existence and effectiveness of a number of independent techniques of the stratospheric aerosol survey. Twilight technique is cost-effective for long-time analysis that is required to study the trends and their possible reasons.

## Acknowledgments

The work is supported by the Russian Foundation for Basic Research, grant 16-05-00170-a.